\documentclass[11pt,twocolumns,article,nofontune]{IEEEtran}

\usepackage{xurl}  
\usepackage[utf8]{inputenc}
\usepackage{xcolor}
\usepackage{amsmath}
\usepackage{amssymb}
\usepackage{xspace}
\usepackage{epsfig}
\usepackage{balance}

\usepackage[acronyms,nonumberlist,nopostdot,nomain,nogroupskip,acronymlists={hidden}]{glossaries}
\newglossary[algh]{hidden}{acrh}{acnh}{Hidden Acronyms}

\usepackage{booktabs}
\usepackage{tabularx}

\usepackage{tikz}
\usepackage{pgfplots}
\pgfplotsset{compat=newest}
\pgfplotsset{plot coordinates/math parser=false}
\newlength\fheight
\newlength\fwidth
\usetikzlibrary{plotmarks,patterns,decorations.pathreplacing,backgrounds,calc,arrows,arrows.meta,spy,matrix,scopes}
\usepgfplotslibrary{patchplots,groupplots}
\usepackage{tikzscale}
\usepackage[draft]{hyperref}

\newif\ifexttikz
\exttikzfalse

\ifexttikz
	\usetikzlibrary{external}
	\usepackage{fontspec}
\fi

\usepackage{multirow}

\usepackage[font=footnotesize]{subcaption}
\usepackage[font=footnotesize]{caption}

\usepackage{mathtools}
\usepackage[numbers,sort&compress]{natbib}

\usepackage{soul}

\newacronym{3gpp}{3GPP}{3rd Generation Partnership Project}
\newacronym{4g}{4G}{4th generation}
\newacronym{5g}{5G}{5th generation}
\newacronym{6g}{6G}{6th generation}
\newacronym{5gc}{5GC}{5G Core}
\newacronym{adc}{ADC}{Analog to Digital Converter}
\newacronym{aerpaw}{AERPAW}{Aerial Experimentation and Research Platform for Advanced Wireless}
\newacronym{ai}{AI}{Artificial Intelligence}
\newacronym{aimd}{AIMD}{Additive Increase Multiplicative Decrease}
\newacronym{am}{AM}{Acknowledged Mode}
\newacronym{amc}{AMC}{Adaptive Modulation and Coding}
\newacronym{amf}{AMF}{Access and Mobility Management Function}
\newacronym{aops}{AOPS}{Adaptive Order Prediction Scheduling}
\newacronym{api}{API}{Application Programming Interface}
\newacronym{apn}{APN}{Access Point Name}
\newacronym{ap}{AP}{Application Protocol}
\newacronym{aqm}{AQM}{Active Queue Management}
\newacronym{ausf}{AUSF}{Authentication Server Function}
\newacronym{avc}{AVC}{Advanced Video Coding}
\newacronym{awgn}{AGWN}{Additive White Gaussian Noise}
\newacronym{balia}{BALIA}{Balanced Link Adaptation Algorithm}
\newacronym{bbu}{BBU}{Base Band Unit}
\newacronym{bdp}{BDP}{Bandwidth-Delay Product}
\newacronym{ber}{BER}{Bit Error Rate}
\newacronym{bf}{BF}{Beamforming}
\newacronym{bler}{BLER}{Block Error Rate}
\newacronym{brr}{BRR}{Bayesian Ridge Regressor}
\newacronym{bs}{BS}{Base Station}
\newacronym{bsr}{BSR}{Buffer Status Report}
\newacronym{bss}{BSS}{Business Support System}
\newacronym{ca}{CA}{Carrier Aggregation}
\newacronym{caas}{CaaS}{Connectivity-as-a-Service}
\newacronym{cb}{CB}{Code Block}
\newacronym{cc}{CC}{Congestion Control}
\newacronym{ccid}{CCID}{Congestion Control ID}
\newacronym{cco}{CC}{Carrier Component}
\newacronym{cdd}{CDD}{Cyclic Delay Diversity}
\newacronym{cdf}{CDF}{Cumulative Distribution Function}
\newacronym{cdn}{CDN}{Content Distribution Network}
\newacronym{cn}{CN}{Core Network}
\newacronym{codel}{CoDel}{Controlled Delay Management}
\newacronym{comac}{COMAC}{Converged Multi-Access and Core}
\newacronym{cord}{CORD}{Central Office Re-architected as a Datacenter}
\newacronym{cornet}{CORNET}{COgnitive Radio NETwork}
\newacronym{cosmos}{COSMOS}{Cloud Enhanced Open Software Defined Mobile Wireless Testbed for City-Scale Deployment}
\newacronym{cots}{COTS}{Commercial Off-the-Shelf}
\newacronym{cp}{CP}{Control Plane}
\newacronym{cyp}{CP}{Cyclic Prefix}
\newacronym{up}{UP}{User Plane}
\newacronym{cpu}{CPU}{Central Processing Unit}
\newacronym{cqi}{CQI}{Channel Quality Information}
\newacronym{cr}{CR}{Cognitive Radio}
\newacronym{cran}{CRAN}{Cloud \gls{ran}}
\newacronym{crs}{CRS}{Cell Reference Signal}
\newacronym{csi}{CSI}{Channel State Information}
\newacronym{csirs}{CSI-RS}{Channel State Information - Reference Signal}
\newacronym{cu}{CU}{Central Unit}
\newacronym{d2tcp}{D$^2$TCP}{Deadline-aware Data center TCP}
\newacronym{d3}{D$^3$}{Deadline-Driven Delivery}
\newacronym{dac}{DAC}{Digital to Analog Converter}
\newacronym{dag}{DAG}{Directed Acyclic Graph}
\newacronym{das}{DAS}{Distributed Antenna System}
\newacronym{dash}{DASH}{Dynamic Adaptive Streaming over HTTP}
\newacronym{dc}{DC}{Dual Connectivity}
\newacronym{dccp}{DCCP}{Datagram Congestion Control Protocol}
\newacronym{dce}{DCE}{Direct Code Execution}
\newacronym{dci}{DCI}{Downlink Control Information}
\newacronym{dctcp}{DCTCP}{Data Center TCP}
\newacronym{dl}{DL}{Deep Learning}
\newacronym{dmr}{DMR}{Deadline Miss Ratio}
\newacronym{dmrs}{DMRS}{DeModulation Reference Signal}
\newacronym{drlcc}{DRL-CC}{Deep Reinforcement Learning Congestion Control}
\newacronym{drs}{DRS}{Discovery Reference Signal}
\newacronym{du}{DU}{Distributed Unit}
\newacronym{e2e}{E2E}{end-to-end}
\newacronym{ecaas}{ECaaS}{Edge-Cloud-as-a-Service}
\newacronym{ecn}{ECN}{Explicit Congestion Notification}
\newacronym{edf}{EDF}{Earliest Deadline First}
\newacronym{embb}{eMBB}{Enhanced Mobile Broadband}
\newacronym{empower}{EMPOWER}{EMpowering transatlantic PlatfOrms for advanced WirEless Research}
\newacronym{enb}{eNB}{evolved Node Base}
\newacronym{endc}{EN-DC}{E-UTRAN-\gls{nr} \gls{dc}}
\newacronym{epc}{EPC}{Evolved Packet Core}
\newacronym{eps}{EPS}{Evolved Packet System}
\newacronym{es}{ES}{Edge Server}
\newacronym{etsi}{ETSI}{European Telecommunications Standards Institute}
\newacronym[firstplural=Estimated Times of Arrival (ETAs)]{eta}{ETA}{Estimated Time of Arrival}
\newacronym{eutran}{E-UTRAN}{Evolved Universal Terrestrial Access Network}
\newacronym{faas}{FaaS}{Function-as-a-Service}
\newacronym{fapi}{FAPI}{Functional Application Platform Interface}
\newacronym{fdd}{FDD}{Frequency Division Duplexing}
\newacronym{fdm}{FDM}{Frequency Division Multiplexing}
\newacronym{fdma}{FDMA}{Frequency Division Multiple Access}
\newacronym{fed4fire}{FED4FIRE+}{Federation 4 Future Internet Research and Experimentation Plus}
\newacronym{fir}{FIR}{Finite Impulse Response}
\newacronym{fit}{FIT}{Future \acrlong{iot}}
\newacronym{fpga}{FPGA}{Field Programmable Gate Array}
\newacronym{fr2}{FR2}{Frequency Range 2}
\newacronym{fscc}{FSCC}{Flow Sharing Congestion Control}
\newacronym{ftp}{FTP}{File Transfer Protocol}
\newacronym{fw}{FW}{Flow Window}
\newacronym{ge}{GE}{Gaussian Elimination}
\newacronym{gnb}{gNB}{Next Generation Node Base}
\newacronym{gop}{GOP}{Group of Pictures}
\newacronym{gpr}{GPR}{Gaussian Process Regressor}
\newacronym{gpu}{GPU}{Graphics Processing Unit}
\newacronym{gtp}{GTP}{GPRS Tunneling Protocol}
\newacronym{gtpc}{GTP-C}{GPRS Tunnelling Protocol Control Plane}
\newacronym{gtpu}{GTP-U}{GPRS Tunnelling Protocol User Plane}
\newacronym{gtpv2c}{GTPv2-C}{\gls{gtp} v2 - Control}
\newacronym{gw}{GW}{Gateway}
\newacronym{harq}{HARQ}{Hybrid Automatic Repeat reQuest}
\newacronym{hetnet}{HetNet}{Heterogeneous Network}
\newacronym{hh}{HH}{Hard Handover}
\newacronym{hol}{HOL}{Head-of-Line}
\newacronym{hqf}{HQF}{Highest-quality-first}
\newacronym{hss}{HSS}{Home Subscription Server}
\newacronym{http}{HTTP}{HyperText Transfer Protocol}
\newacronym{ia}{IA}{Initial Access}
\newacronym{iab}{IAB}{Integrated Access and Backhaul}
\newacronym{ic}{IC}{Incident Command}
\newacronym{ietf}{IETF}{Internet Engineering Task Force}
\newacronym{ims}{IMS}{Infrastructure Management Service}
\newacronym{imsi}{IMSI}{International Mobile Subscriber Identity}
\newacronym{imt}{IMT}{International Mobile Telecommunication}
\newacronym{iot}{IoT}{Internet of Things}
\newacronym{ip}{IP}{Internet Protocol}
\newacronym{itu}{ITU}{International Telecommunication Union}
\newacronym{kpi}{KPI}{Key Performance Indicator}
\newacronym{kpm}{KPM}{Key Performance Measurement}
\newacronym{kvm}{KVM}{Kernel-based Virtual Machine}
\newacronym{los}{LOS}{Line-of-Sight}
\newacronym{lsm}{LSM}{Link-to-System Mapping}
\newacronym{lstm}{LSTM}{Long Short Term Memory}
\newacronym{lte}{LTE}{Long Term Evolution}
\newacronym{lxc}{LXC}{Linux Container}
\newacronym{m2m}{M2M}{Machine to Machine}
\newacronym{mac}{MAC}{Medium Access Control}
\newacronym{manet}{MANET}{Mobile Ad Hoc Network}
\newacronym{mano}{MANO}{Management and Orchestration}
\newacronym{mc}{MC}{Multi-Connectivity}
\newacronym{mcc}{MCC}{Mobile Cloud Computing}
\newacronym{mchem}{MCHEM}{Massive Channel Emulator}
\newacronym{mcs}{MCS}{Modulation and Coding Scheme}
\newacronym{mec}{MEC}{Multi-access Edge Computing}
\newacronym{mec2}{MEC}{Mobile Edge Cloud}
\newacronym{mfc}{MFC}{Mobile Fog Computing}
\newacronym{mgen}{MGEN}{Multi-Generator}
\newacronym{mi}{MI}{Mutual Information}
\newacronym{mib}{MIB}{Master Information Block}
\newacronym{miesm}{MIESM}{Mutual Information Based Effective SINR}
\newacronym{mimo}{MIMO}{Multiple Input, Multiple Output}
\newacronym{ml}{ML}{Machine Learning}
\newacronym{mlr}{MLR}{Maximum-local-rate}
\newacronym[plural=\gls{mme}s,firstplural=Mobility Management Entities (MMEs)]{mme}{MME}{Mobility Management Entity}
\newacronym{mmtc}{mMTC}{Massive Machine-Type Communications}
\newacronym{mmwave}{mmWave}{millimeter wave}
\newacronym{mpdccp}{MP-DCCP}{Multipath Datagram Congestion Control Protocol}
\newacronym{mptcp}{MPTCP}{Multipath TCP}
\newacronym{mr}{MR}{Maximum Rate}
\newacronym{mrdc}{MR-DC}{Multi \gls{rat} \gls{dc}}
\newacronym{mse}{MSE}{Mean Square Error}
\newacronym{mss}{MSS}{Maximum Segment Size}
\newacronym{mt}{MT}{Mobile Termination}
\newacronym{mtd}{MTD}{Machine-Type Device}
\newacronym{mtu}{MTU}{Maximum Transmission Unit}
\newacronym{mumimo}{MU-MIMO}{Multi-user \gls{mimo}}
\newacronym{mvno}{MVNO}{Mobile Virtual Network Operator}
\newacronym{nalu}{NALU}{Network Abstraction Layer Unit}
\newacronym{nas}{NAS}{Non-Access Stratum}
\newacronym{nbiot}{NB-IoT}{Narrow Band IoT}
\newacronym{nfv}{NFV}{Network Function Virtualization}
\newacronym{nfvi}{NFVI}{Network Function Virtualization Infrastructure}
\newacronym{ni}{NI}{Network Interfaces}
\newacronym{nic}{NIC}{Network Interface Card}
\newacronym{nlos}{NLOS}{Non-Line-of-Sight}
\newacronym{now}{NOW}{Non Overlapping Window}
\newacronym{nsm}{NSM}{Network Service Mesh}
\newacronym[type=hidden]{nr}{NR}{New Radio}
\newacronym{nrf}{NRF}{Network Repository Function}
\newacronym{nsa}{NSA}{Non Stand Alone}
\newacronym{nse}{NSE}{Network Slicing Engine}
\newacronym{nssf}{NSSF}{Network Slice Selection Function}
\newacronym{o2i}{O2I}{Outdoor to Indoor}
\newacronym{oai}{OAI}{OpenAirInterface}
\newacronym{oaicn}{OAI-CN}{\gls{oai} \acrlong{cn}}
\newacronym{oairan}{OAI-RAN}{\acrlong{oai} \acrlong{ran}}
\newacronym{oam}{OAM}{Operations, Administration and Maintenance}
\newacronym{ofdm}{OFDM}{Orthogonal Frequency Division Multiplexing}
\newacronym{olia}{OLIA}{Opportunistic Linked Increase Algorithm}
\newacronym{omec}{OMEC}{Open Mobile Evolved Core}
\newacronym{onap}{ONAP}{Open Network Automation Platform}
\newacronym{onf}{ONF}{Open Networking Foundation}
\newacronym{onos}{ONOS}{Open Networking Operating System}
\newacronym{oom}{OOM}{\gls{onap} Operations Manager}
\newacronym{opnfv}{OPNFV}{Open Platform for \gls{nfv}}
\newacronym[type=hidden]{oran}{O-RAN}{O-RAN}
\newacronym{orbit}{ORBIT}{Open-Access Research Testbed for Next-Generation Wireless Networks}
\newacronym{os}{OS}{Operating System}
\newacronym{oss}{OSS}{Operations Support System}
\newacronym{pa}{PA}{Position-aware}
\newacronym{pase}{PASE}{Prioritization, Arbitration, and Self-adjusting Endpoints}
\newacronym{pawr}{PAWR}{Platforms for Advanced Wireless Research}
\newacronym{pbch}{PBCH}{Physical Broadcast Channel}
\newacronym{pcef}{PCEF}{Policy and Charging Enforcement Function}
\newacronym{pcfich}{PCFICH}{Physical Control Format Indicator Channel}
\newacronym{pcrf}{PCRF}{Policy and Charging Rules Function}
\newacronym{pdcch}{PDCCH}{Physical Downlink Control Channel}
\newacronym{pdcp}{PDCP}{Packet Data Convergence Protocol}
\newacronym{pdsch}{PDSCH}{Physical Downlink Shared Channel}
\newacronym{pdu}{PDU}{Packet Data Unit}
\newacronym{pf}{PF}{Proportional Fair}
\newacronym{pgw}{PGW}{Packet Gateway}
\newacronym{phich}{PHICH}{Physical Hybrid ARQ Indicator Channel}
\newacronym{phy}{PHY}{Physical}
\newacronym{pmch}{PMCH}{Physical Multicast Channel}
\newacronym{pmi}{PMI}{Precoding Matrix Indicators}
\newacronym{powder}{POWDER}{Platform for Open Wireless Data-driven Experimental Research}
\newacronym{ppo}{PPO}{Proximal Policy Optimization}
\newacronym{ppp}{PPP}{Poisson Point Process}
\newacronym{prach}{PRACH}{Physical Random Access Channel}
\newacronym{prb}{PRB}{Physical Resource Block}
\newacronym{psnr}{PSNR}{Peak Signal to Noise Ratio}
\newacronym{pss}{PSS}{Primary Synchronization Signal}
\newacronym{pucch}{PUCCH}{Physical Uplink Control Channel}
\newacronym{pusch}{PUSCH}{Physical Uplink Shared Channel}
\newacronym{qam}{QAM}{Quadrature Amplitude Modulation}
\newacronym{qci}{QCI}{\gls{qos} Class Identifier}
\newacronym{qoe}{QoE}{Quality of Experience}
\newacronym{qos}{QoS}{Quality of Service}
\newacronym{quic}{QUIC}{Quick UDP Internet Connections}
\newacronym{rach}{RACH}{Random Access Channel}
\newacronym{ran}{RAN}{Radio Access Network}
\newacronym[firstplural=Radio Access Technologies (RATs)]{rat}{RAT}{Radio Access Technology}
\newacronym{rcn}{RCN}{Research Coordination Network}
\newacronym{rc}{RC}{RAN Control}
\newacronym{rec}{REC}{Radio Edge Cloud}
\newacronym{red}{RED}{Random Early Detection}
\newacronym{renew}{RENEW}{Reconfigurable Eco-system for Next-generation End-to-end Wireless}
\newacronym{rf}{RF}{Radio Frequency}
\newacronym{rfc}{RFC}{Request for Comments}
\newacronym{rfr}{RFR}{Random Forest Regressor}
\newacronym{ric}{RIC}{RAN Intelligent Controller}
\newacronym{rlc}{RLC}{Radio Link Control}
\newacronym{rlf}{RLF}{Radio Link Failure}
\newacronym{rlnc}{RLNC}{Random Linear Network Coding}
\newacronym{rmr}{RMR}{RIC Message Router}
\newacronym{rmse}{RMSE}{Root Mean Squared Error}
\newacronym{rnis}{RNIS}{Radio Network Information Service}
\newacronym{rr}{RR}{Round Robin}
\newacronym{rrc}{RRC}{Radio Resource Control}
\newacronym{rrm}{RRM}{Radio Resource Management}
\newacronym{rru}{RRU}{Remote Radio Unit}
\newacronym{rs}{RS}{Remote Server}
\newacronym{rsrp}{RSRP}{Reference Signal Received Power}
\newacronym{rsrq}{RSRQ}{Reference Signal Received Quality}
\newacronym{rss}{RSS}{Received Signal Strength}
\newacronym{rssi}{RSSI}{Received Signal Strength Indicator}
\newacronym{rtt}{RTT}{Round Trip Time}
\newacronym{ru}{RU}{Radio Unit}
\newacronym{rw}{RW}{Receive Window}
\newacronym{rx}{RX}{Receiver}
\newacronym{s1ap}{S1AP}{S1 Application Protocol}
\newacronym{sa}{SA}{standalone}
\newacronym{sack}{SACK}{Selective Acknowledgment}
\newacronym{sap}{SAP}{Service Access Point}
\newacronym{sc2}{SC2}{Spectrum Collaboration Challenge}
\newacronym{scef}{SCEF}{Service Capability Exposure Function}
\newacronym{sch}{SCH}{Secondary Cell Handover}
\newacronym{scoot}{SCOOT}{Split Cycle Offset Optimization Technique}
\newacronym{sctp}{SCTP}{Stream Control Transmission Protocol}
\newacronym{sdap}{SDAP}{Service Data Adaptation Protocol}
\newacronym{sdk}{SDK}{Software Development Kit}
\newacronym{sdm}{SDM}{Space Division Multiplexing}
\newacronym{sdma}{SDMA}{Spatial Division Multiple Access}
\newacronym{sdn}{SDN}{Software-defined Networking}
\newacronym{sdr}{SDR}{Software-defined Radio}
\newacronym{seba}{SEBA}{SDN-Enabled Broadband Access}
\newacronym{sgsn}{SGSN}{Serving GPRS Support Node}
\newacronym{sgw}{SGW}{Service Gateway}
\newacronym{si}{SI}{Study Item}
\newacronym{sib}{SIB}{Secondary Information Block}
\newacronym{sinr}{SINR}{Signal to Interference plus Noise Ratio}
\newacronym{sip}{SIP}{Session Initiation Protocol}
\newacronym{siso}{SISO}{Single Input, Single Output}
\newacronym{sla}{SLA}{Service Level Agreement}
\newacronym{sm}{SM}{Service Model}
\newacronym{smf}{SMF}{Session Management Function}
\newacronym{smo}{SMO}{Service Management and Orchestration}
\newacronym{sms}{SMS}{Short Message Service}
\newacronym{smsgmsc}{SMS-GMSC}{\gls{sms}-Gateway}
\newacronym{snr}{SNR}{Signal-to-Noise-Ratio}
\newacronym{son}{SON}{Self-Organizing Network}
\newacronym{sptcp}{SPTCP}{Single Path TCP}
\newacronym{srb}{SRB}{Service Radio Bearer}
\newacronym{srn}{SRN}{Standard Radio Node}
\newacronym{srs}{SRS}{Sounding Reference Signal}
\newacronym{ss}{SS}{Synchronization Signal}
\newacronym{sss}{SSS}{Secondary Synchronization Signal}
\newacronym{st}{ST}{Spanning Tree}
\newacronym{svc}{SVC}{Scalable Video Coding}
\newacronym{tb}{TB}{Transport Block}
\newacronym{tcp}{TCP}{Transmission Control Protocol}
\newacronym{tdd}{TDD}{Time Division Duplexing}
\newacronym{tdm}{TDM}{Time Division Multiplexing}
\newacronym{tdma}{TDMA}{Time Division Multiple Access}
\newacronym{tfl}{TfL}{Transport for London}
\newacronym{tfrc}{TFRC}{TCP-Friendly Rate Control}
\newacronym{tft}{TFT}{Traffic Flow Template}
\newacronym{tgen}{TGEN}{Traffic Generator}
\newacronym{tip}{TIP}{Telecom Infra Project}
\newacronym{tm}{TM}{Transparent Mode}
\newacronym{to}{TO}{Telco Operator}
\newacronym{tr}{TR}{Technical Report}
\newacronym{trp}{TRP}{Transmitter Receiver Pair}
\newacronym{ts}{TS}{Technical Specification}
\newacronym{tti}{TTI}{Transmission Time Interval}
\newacronym{ttt}{TTT}{Time-to-Trigger}
\newacronym{tx}{TX}{Transmitter}
\newacronym{uas}{UAS}{Unmanned Aerial System}
\newacronym{uav}{UAV}{Unmanned Aerial Vehicle}
\newacronym{udm}{UDM}{Unified Data Management}
\newacronym{udp}{UDP}{User Datagram Protocol}
\newacronym{udr}{UDR}{Unified Data Repository}
\newacronym{ue}{UE}{User Equipment}
\newacronym{uhd}{UHD}{\gls{usrp} Hardware Driver}
\newacronym{ul}{UL}{Uplink}
\newacronym{um}{UM}{Unacknowledged Mode}
\newacronym{uml}{UML}{Unified Modeling Language}
\newacronym{upa}{UPA}{Uniform Planar Array}
\newacronym{upf}{UPF}{User Plane Function}
\newacronym{urllc}{URLLC}{Ultra Reliable and Low Latency Communications}
\newacronym{usa}{U.S.}{United States}
\newacronym{usim}{USIM}{Universal Subscriber Identity Module}
\newacronym{usrp}{USRP}{Universal Software Radio Peripheral}
\newacronym{utc}{UTC}{Urban Traffic Control}
\newacronym{vim}{VIM}{Virtualization Infrastructure Manager}
\newacronym{vm}{VM}{Virtual Machine}
\newacronym{vnf}{VNF}{Virtual Network Function}
\newacronym{volte}{VoLTE}{Voice over \gls{lte}}
\newacronym{voltha}{VOLTHA}{Virtual OLT HArdware Abstraction}
\newacronym{vr}{VR}{Virtual Reality}
\newacronym{vran}{vRAN}{Virtualized \gls{ran}}
\newacronym{vss}{VSS}{Video Streaming Server}
\newacronym{wbf}{WBF}{Wired Bias Function}
\newacronym{wf}{WF}{Waterfilling}
\newacronym{wg}{WG}{Working Group}
\newacronym{wlan}{WLAN}{Wireless Local Area Network}
\newacronym{osm}{OSM}{Open Source \gls{nfv} Management and Orchestration}
\newacronym{pnf}{PNF}{Physical Network Function}
\newacronym{drl}{DRL}{Deep Reinforcement Learning}
\newacronym{mtc}{MTC}{Machine-type Communications}
\newacronym{osc}{OSC}{O-RAN Software Community}
\newacronym{mns}{MnS}{Management Services}
\newacronym{ves}{VES}{\gls{vnf} Event Stream}
\newacronym{ei}{EI}{Enrichment Information}
\newacronym{fh}{FH}{Fronthaul}
\newacronym{fft}{FFT}{Fast Fourier Transform}
\newacronym{laa}{LAA}{Licensed-Assisted Access}
\newacronym{plfs}{PLFS}{Physical Layer Frequency Signals}
\newacronym{ptp}{PTP}{Precision Time Protocol}
\newacronym{cnn}{CNN}{Convolutional Neural Network}
\newacronym{aoa}{AoA}{Angle of Arrival}
\newacronym{xr}{XR}{Extended Reality}
\newacronym{icc}{ICC}{Intelligence Coordination Controller}
\newacronym{smos}{SMOS}{SMO Services}
\newacronym{focom}{FOCOM}{Federated O-Cloud Orchestration and Management}
\newacronym{nfo}{NFO}{Network Function Orchestration}
\newacronym{dms}{DMS}{Deployment Management Service}
\newacronym{llm}{LLM}{Large Language Model}
\newacronym{isac}{ISAC}{Integrated Sensing and Communications}
\newacronym{mig}{MIG}{Multi-Instance GPU}
\newacronym{dsp}{DSP}{Digital Signal Processing}
\newacronym{sic}{SIC}{Spectrum Intelligent Controller}
\newacronym{rfsoc}{RFSoC}{Radio Frequency System on Chip}
\newacronym{ntn}{NTN}{Non-Terrestrial Network}
\newacronym{rfi}{RFI}{Radio-Frequency Interference}
\newacronym{pal}{PAL}{Priority Access License}
\newacronym{gaa}{GAA}{General Authorized Access}
\newacronym{ct}{CT}{Continuous Testing}

\newacronym{wrc}{WRC}{World Radiocommunication Conference}
\newacronym{xpr}{XPR}{Cross Polarization Ratio}
\newacronym{sthz}{Sub-THz}{sub-terahertz}
\newacronym{thz}{THz}{terahertz}
\newacronym{aclr}{ACLR}{Adjacent Channel Leakage power Ratio}
\newacronym{fr}{FR}{Frequency Range}
\newacronym{mbs}{MBS}{Multicast and Broadcast Service}
\newacronym{mgws}{MGWS}{Multi-Gigabit Wireless Systems}
\newacronym{uwb}{UWB}{Ultra-Wideband}
\newacronym{mvdss}{MVDDS}{Multi-Channel Video and Data Distribution Service}
\newacronym{dbs}{DBS}{Direct Broadcast Satellite}
\newacronym{rls}{RLS}{Radiolocation Service}
\newacronym{ms}{MS}{Mobile Service}
\newacronym{sr}{SR}{Space Research Service}
\newacronym{ra}{RA}{Radio Astronomy Service}

\newacronym{eess}{EESS}{Earth Exploration-Satellite Service}
\newacronym{faa}{FAA}{Federal Aviation Administration}
\newacronym{fcc}{FCC}{Federal Communications Commission}

\newacronym{metsat}{MetSat}{Meteorological Satellite Service}
\newacronym{is}{IS}{Inter-satellite Service}
\newacronym{fs}{FS}{Fixed Service}
\newacronym{fss}{FSS}{Fixed Satellite Service}
\newacronym{em}{EM}{Electro Magnetic}
\newacronym{lna}{LNA}{Low-Noise Amplifier}
\newacronym{dss}{DSS}{Dynamic Spectrum Sharing}
\newacronym{pdf}{PDF}{Probability Density Function}
\newacronym{mpc}{MPC}{Multipath Component}
\newacronym{cbrs}{CBRS}{Citizen Broadband Radio Service}
\newacronym{sApp}{sApp}{Spectrum Application}
\newacronym{dsa}{DSA}{Dynamic Spectrum Access}
\newacronym{sas}{SAS}{Spectrum Access System}
\newacronym{dt}{DT}{Digital Twin}
\newacronym{rt}{RT}{Ray Tracing}

\newacronym{lsa}{LSA}{Licensed Shared Access}
\newacronym{papr}{PAPR}{Peak-to-Average Power Ratio}
\newacronym{afc}{AFC}{Automated Frequency Coordination}
\glsdisablehyper

\newcommand{\projName}{Open Spectrum\xspace}

\definecolor{desireRed}{RGB}{230,57,60}%
\definecolor{darkPurple}{RGB}{59,31,43}%
\definecolor{springGreen}{RGB}{37,223,145}%
\definecolor{queenBlue}{RGB}{69,123,157}%
\definecolor{spaceCadet}{RGB}{29,53,87}%

\usepgfplotslibrary{colormaps} 

\usepackage{dblfloatfix}

\IEEEoverridecommandlockouts

\definecolor{svcSensing}  {rgb}{0.0000, 0.4470, 0.7410} 
\definecolor{svcRadNav}   {rgb}{0.8500, 0.3250, 0.0980} 
\definecolor{svcRadiolo}  {rgb}{0.9000, 0.7500, 0.1000} 
\definecolor{svcCellular}{rgb}{0,0,0} 

\pgfplotsset{
	cdfcurve/.style={
		line width=0.8pt,
		mark size=2pt,
		mark repeat=30,
		mark phase=15,
	},
}

\pgfplotsset{colormap={turbo}{%
		rgb255=(48,18,59)   rgb255=(70,107,227) rgb255=(35,176,221)%
		rgb255=(54,217,164) rgb255=(149,233,49) rgb255=(244,198,33)%
		rgb255=(243,118,33) rgb255=(184,32,2)   rgb255=(122,4,3)}}

\begin{document}

\title{From Open RAN to Open Spectrum:\\A Programmable, Intelligent Architecture for Multi-Service Spectrum Coexistence}

\author{Michele~Polese,~\IEEEmembership{Senior Member,~IEEE,}
	Minh~Dat~Nguyen,~\IEEEmembership{Member,~IEEE,}
	Paolo~Testolina,~\IEEEmembership{Member,~IEEE,}
	and~Tommaso~Melodia,~\IEEEmembership{Fellow,~IEEE}%
	\thanks{M.~Polese, M.~D.~Nguyen, P.~Testolina, and T.~Melodia are with the Institute for Intelligent Networked Systems, Northeastern University, Boston, MA 02115, USA (e-mail: \{m.polese, minhd.nguyen, p.testolina, t.melodia\}@northeastern.edu).}
	\thanks{This work was partially supported by the U.S. NSF under award CNS-2434081, by the U.S. Government under Other Transaction number
			W15QKN-21-9-5599 between the National Spectrum Consortium (NSC) and the
			Government, and by OUSW(R\&E) through Army Research Laboratory
Cooperative Agreement Number W911NF-24-2-0065. The U.S. Government is authorized to reproduce and distribute reprints
			for Governmental purposes notwithstanding any copyright notation herein. The views and conclusions contained herein are those of the authors and should not
			be interpreted as necessarily representing the official policies or endorsements, either
			expressed or implied, of the U.S. Government.}
            \thanks{DISTRIBUTION STATEMENT A. APPROVED FOR PUBLIC RELEASE; DISTRIBUTION IS UNLIMITED}%
}

\makeatletter
\patchcmd{\@maketitle}
  {\addvspace{0.5\baselineskip}\egroup}
  {\addvspace{-1.75\baselineskip}\egroup}
  {}
  {}
\makeatother

\maketitle

\begin{abstract}

Considering sharing or coexistence from the perspective of spectrum alone fails to recognize that any spectrum-enabled service also requires (i) radio and processing infrastructure and (ii) a protocol stack, including waveforms and signal processing pipelines. The efficiency of spectrum coexistence frameworks such as \gls{cbrs} is thus limited to optimizing resource allocation across a single dimension. How to address this limitation, however, remains an open challenge, especially considering the diversity of requirements and operational modes across spectrum services (e.g., sensing, communications, navigation, or positioning).

This article introduces \projName, an architecture that brings softwarization, programmability, and open interfaces to heterogeneous spectrum services, extending the open \gls{ran} principles beyond wireless networking. We propose to combine spectrum, services, and infrastructure in a common pool. Its resources are shared and orchestrated by a \gls{sic}, with plug-and-play spectrum applications, i.e., \glspl{sApp}, and data-driven \gls{rfi} modeling using \glspl{dt}.
We describe the \projName architecture, shared infrastructure pool, and operational workflows for tenant onboarding and incentives, conflict resolution, and service sharing across sensing, radionavigation, radiolocation, and cellular systems.
System-level simulations using the BostonTwin urban \gls{dt} and Sionna ray tracing show that there exist performance-driven incentives in sharing infrastructure and sharing across multiple services, enabling increased access to spectrum and improvement in median \gls{sinr} of up to $12$ dB. 
\end{abstract}


\glsresetall
\glsunset{oran}


\vspace{-2mm}
\section{Introduction}
\label{sec:intro}

Exclusive spectrum licensing, where a regulator assigns a frequency band to a single service or operator, who builds dedicated infrastructure to use it, is increasingly inadequate.
The rise of \gls{isac}, high-precision positioning, space networking, and sensing creates demand for spectrum resources that static allocation cannot accommodate~\cite{SS:Inter-Tech19}. 
At the same time, unlicensed access does not provide guarantees to services with sensitivity to \gls{rfi} (e.g., passive Earth exploration, radioastronomy, and sensing).
Access requirements are also dynamic and shift over space and time:
radioastronomy observations, for instance, benefit from bands outside established allocations due to Doppler shifts from moving targets.


Regulatory and industry innovations have begun to address this.
Frameworks such as \gls{cbrs}, \gls{dsa}, 
and \gls{afc}
enable dynamic frequency sharing, albeit with a fundamental limitation: they address spectrum in isolation, without considering the infrastructure and services that depend on it.
A \gls{cbrs} \gls{sas} can grant frequency access, but it cannot help a radar operator who lacks the dense infrastructure to deploy a new sensing service, nor can it coordinate the joint use of radio hardware across service types.

\begin{figure}[t!]
	\centering
	\includegraphics[width=\columnwidth]{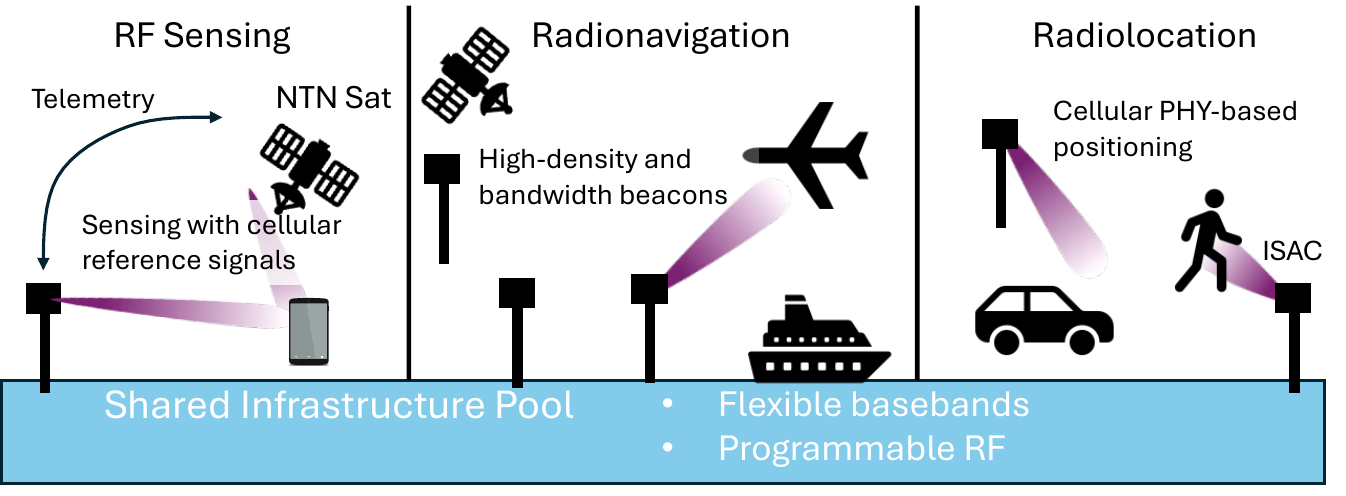}
	\caption{Use cases for spectrum coexistence with heterogeneous services based on a shared infrastructure, services, and spectrum pool.}
	\label{fig:service_sharing}
	\vspace{-5mm}
\end{figure}

This article introduces \projName, an architecture that extends the common-pool resource perspective on spectrum governance~\cite{weiss2017how} to infrastructure and waveform sharing. 
In the common pool, resources are available to users of a heterogeneous set of services (tenants, as shown in Fig.~\ref{fig:service_sharing}). They are willing to share one or more elements among
spectrum allocations, which can be based on legacy systems, and radio infrastructure; and to provide their service on a shared portion of the spectrum. 
%
This creates economic incentives for participation: sparse services (sensing, radar, navigation) gain access to dense infrastructure they could not economically deploy on their own, while telecom operators can dynamically extend their spectrum footprint for connectivity.



The \projName system architecture provides {\em observability} as well as {\em programmability} through a closed control loop that (i) exposes spectrum and infrastructure telemetry; (ii) generates service-level policies beyond \gls{rfi} alone; and (iii) enforces allocations on a programmable radio infrastructure.
A shared infrastructure enables granular conflict resolution across heterogeneous services and, beyond spectrum, allows the same hardware and programmable basebands to deliver multiple spectrum services simultaneously (\emph{sharing services}).


In the remainder of this article, we compare \projName with prior literature and detail the proposed architecture. We then evaluate the opportunities associated with spectrum and infrastructure sharing through a \gls{dt} framework that combines large-scale \gls{rt} with Sionna~\cite{hoydis2023sionnart} with a real-world urban 3D model and multi-service deployments footprints~\cite{testolina2024bostontwin}. 
Our results show the benefits of joint spectrum and infrastructure sharing. Services with cellular-like \gls{rf} parameters gain up to $12$\,dB in median \gls{sinr}. 
The orders-of-magnitude duty-cycle gap between sparse and cellular services leaves substantial idle time-frequency resources, which \projName reclaims through coordinated scheduling without degrading incumbent \gls{qos}.

\begin{figure*}[t!]
	\centering
	\includegraphics[width=0.8\textwidth]{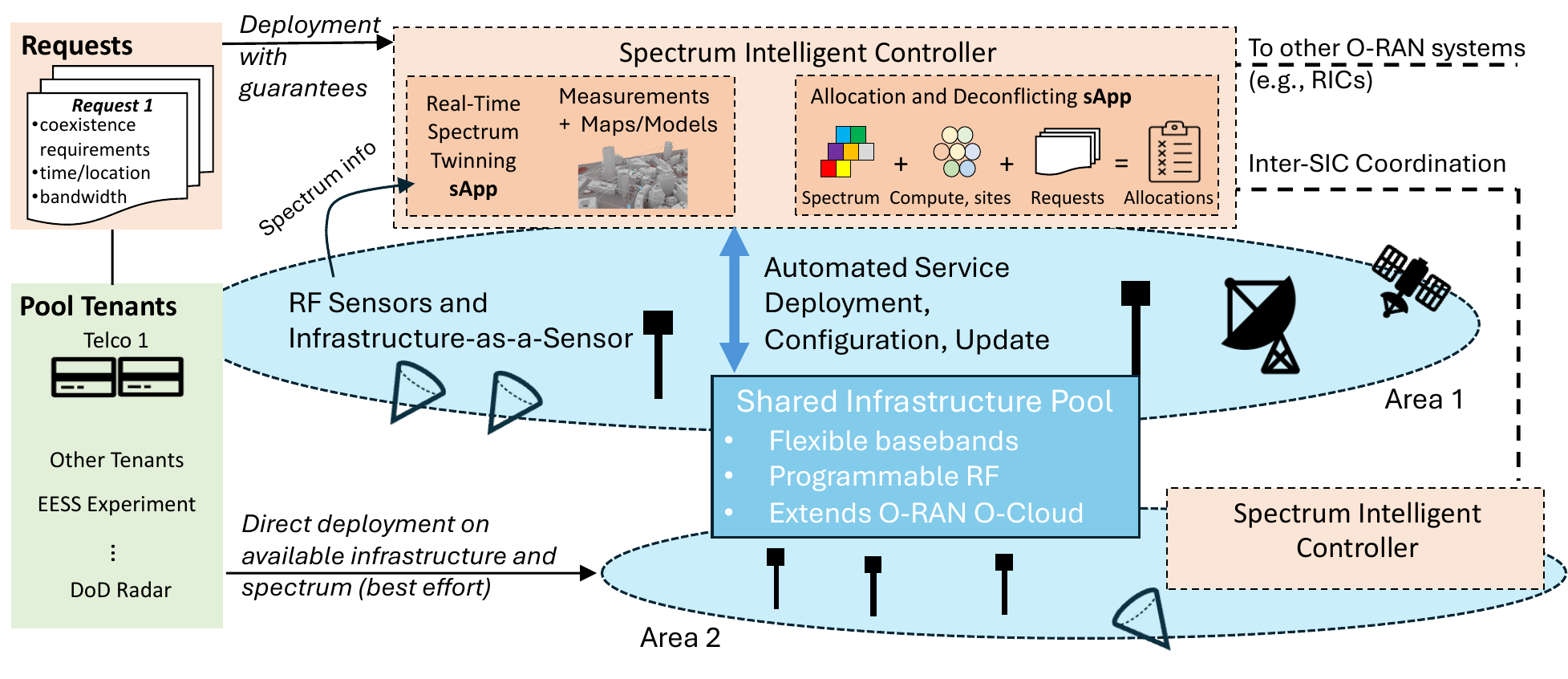}
	\caption{\projName architecture: the SIC extends the \gls{oran} RIC to coordinate heterogeneous services via sApps, with DT-based \gls{rfi} modeling and a shared infrastructure pool.}
	\label{fig:architecture}
	\vspace{-.45cm}
\end{figure*}


\vspace{-2mm}
\section{Related Work}
\label{sec:related}

We identify three gaps in prior work that \projName addresses.

\textbf{Pairwise coexistence without multi-service coordination.}
Most spectrum sharing research examines coexistence between two system types: 
sensing-communication pairs above 100~GHz~\cite{SS:100GHz23}, passive sensing interference~\cite{testolina2024modeling}, and \gls{oran}-based \gls{cbrs} coexistence~\cite{acharya2023mitra}.
Multi-service works in the upper midband~\cite{SS:UpperMid24} address band-specific allocation but do not provide a unified cross-service controller or joint infrastructure sharing.
Current database-driven systems (\gls{cbrs}~\cite{SS:cbrs24}, \gls{dsa}~\cite{SS:NTIAAdDynamic24}) 
make largely grant-or-deny spectrum decisions with limited \gls{qos} awareness, leaving sharing beyond spectrum 
unaddressed.


\textbf{Spectrum sharing without infrastructure or service sharing.}
Operators leverage \gls{dss} for flexible 4G/5G carrier allocation and neutral hosting~\cite{bonati2024neutran} for shared cellular infrastructure.
Surveys have catalogued sharing techniques across 5G bands~\cite{SS:Inter-Tech19}, 
and spectrum governance has been studied through common-pool resource theory~\cite{weiss2017how}. 
This identifies boundary rules, proportional allocation, and conflict resolution as prerequisites for sustainable sharing. \projName operationalizes these elements across heterogeneous services, unlike any existing spectrum architecture, which shares spectrum but not infrastructure or services.



\textbf{Service-specific resource models prevent cross-service optimization.} Existing spectrum sharing frameworks define resources using service-specific abstractions: \gls{cbrs} relies on Priority Access Licenses and General Authorized Access tiers~\cite{SS:cbrs24}, 
\gls{dss} operates over \gls{lte}/\gls{nr} resource blocks~\cite{SS:Inter-Tech19}, and satellite-terrestrial coordination relies on geometric exclusion zones~\cite{SS:100GHz23}. 
These representations are fundamentally incompatible: they encode different notions of interference, time-frequency granularity, and protection criteria, preventing unified optimization across communication, sensing, navigation, and radiolocation under a common resource abstraction.

\vspace{-2mm}
\section{\projName Architecture}
\label{sec:arch}

The \projName architecture enables joint sharing of spectrum, services, and infrastructure through three interconnected components: the \gls{sic}, a shared infrastructure pool, and a spectrum monitoring framework.
Figure~\ref{fig:architecture} illustrates the overall architecture. Pool tenants request access to the \gls{sic}, specifying coexistence or frequency requirements, time, bandwidth, and other relevant parameters.
The \gls{sic} processes such requests to provide access \emph{with guarantees}, i.e., satisfying the coexistence or \gls{qos} requirements for the tenant.
\glspl{sApp} plug into the \gls{sic} to implement modular spectrum policies. 
The shared infrastructure pool (bottom) provides the physical resources---compute, radio hardware, and sensors---that services access through the \gls{sic}'s coordination. Tenants can also submit \emph{best effort} service requests to the infrastructure, without guarantees that, if the request is accepted, the service's \gls{rfi} requirements are satisfied.
A \gls{dt} module feeds site-specific propagation intelligence into the \gls{sic}'s decision engine.

\subsection{Spectrum Intelligence Controller}
\vspace{-1mm}

The \gls{sic} is the central coordination entity of \projName, enabling spectrum coordination for heterogeneous service types. It extends the closed-loop control introduced in O-RAN by the \gls{ric} to control and data that have no representation in the \gls{oran} \gls{ric}, e.g., radar pulse schedules, navigation beacon timing, and sensing duty cycles.
\glspl{sic} operate on timescales of milliseconds (conflict detection) to hours (scheduled pool allocation). A \gls{sic} can also interface with O-RAN components, e.g., the Near-RT \gls{ric}, to enforce policies that require tuning or configuration of the cellular \gls{ran} beyond spectrum, compute, and radio resource allocation, e.g., cellular user load balancing or handover.

Multiple \glspl{sic} coordinate across geographic areas in a peer-to-peer fashion, enabling regional spectrum policies (e.g., urban vs.\ coastal) while maintaining local autonomy under a common national framework.
Each \gls{sic} maintains a local state database of active tenants, their current grants, and the \gls{dt}'s propagation cache for its coverage area.
When a sharing request involves a boundary region between two \glspl{sic}, an exchange is coordinated across the two \glspl{sic}, which evaluate shared priority of spectrum and infrastructure access for the stakeholders submitting conflicting requests. Compared to a hierarchical approach, this enables scalability and low latency for cross-boundary interactions.


\textbf{Spectrum Applications (\glspl{sApp}).}
The \gls{sic} supports plug-and-play customization through \glspl{sApp}---modular policy functions that third parties (regulators, service providers, researchers) can deploy and update independently.
For example, a \textit{duty-cycle enforcement sApp} ingests each tenant's activity profile ($\delta_i$, frequency band, geographic area) and verifies that duty-cycle budgets satisfy $\delta_X + \delta_Y \leq 1$ for every co-channel pair---a necessary condition for interference-free operation when the \gls{sic} schedules non-overlapping time slots with appropriate guard intervals.
If violated, it invokes a priority table to determine which tenant's duty cycle to reduce and issues updated grants.
Conflicts between sApps are resolved by the \gls{sic}'s arbitration layer with a configurable precedence order.


A key proposed capability is the integration of a \gls{dt} for data-driven \gls{rfi} modeling~\cite{hoydis2023sionnart, testolina2024bostontwin}.
Rather than relying on static exclusion zones as in current \gls{sas} implementations, the \gls{sic} would use a site-specific \gls{dt} that pre-computes propagation maps and predicts \gls{sinr} at affected receiver locations.
This can enable \gls{qos}-aware allocation decisions that account for actual building geometry, material properties, and multi-path propagation, moving beyond conservative worst-case thresholds.
The evaluation in Section~\ref{sec:eval} uses this \gls{dt} framework for offline characterization; closing the real-time feedback loop between the \gls{dt} and \gls{sic} is a subject of ongoing work.

\vspace{-3mm}
\subsection{Shared Infrastructure Pool}
\vspace{-1mm}

\projName extends the \gls{oran} O-Cloud concept into a shared infrastructure pool.
The rationale is that sparse services (sensing, radar, navigation) lack the economic scale to build dense deployments, while cellular operators have (i) excess infrastructure capacity during off-peak hours and (ii) a need for additional spectrum during peak hours.
Joint sharing creates value for both sides.
The pool includes three resource categories:
\begin{itemize}
\item \textbf{Compute}: Programmable servers hosting virtualized, software-based signal-processing functions, either for \gls{ran}, sensing, or other spectrum users. An automation layer~\cite{bonati2024neutran} manages the service deployment, configuration, and life cycle.
\item \textbf{Radio hardware}: Radio units, software-defined radios, and reconfigurable hardware that can be dynamically reassigned, e.g., wideband \gls{sdr} platforms at cellular sites configured for radar waveform generation during low-traffic periods.
\item \textbf{RF sensors}: Distributed spectrum monitoring devices for regulatory compliance and real-time conflict detection.
\end{itemize}

The pool operates in two modes.
In \textit{scheduled mode}, resources are reserved in advance for services with predictable requirements, e.g., radar pulse schedules with known dwell times and revisit intervals.
The \gls{sic} pre-computes a conflict-free time-frequency plan and distributes it to all affected tenants before the scheduling epoch begins.
In \textit{best-effort mode}, remaining capacity is dynamically allocated to services that tolerate variable access, such as \gls{iot} sensors performing periodic environmental measurements that can defer transmissions by seconds without impact.

\vspace{-3mm}
\subsection{Spectrum Monitoring}
\vspace{-1mm}

\projName leverages two complementary monitoring approaches: dedicated spectrum sensors providing ground-truth measurements of spectral occupancy
analogous to the \gls{cbrs} Environmental Sensing Capability (ESC)~\cite{winnf_esc}, 
and an ``infrastructure-as-sensor'' model where cellular base stations, radar receivers, and navigation beacons continuously report received signal characteristics to the \gls{sic}~\cite{lacava2025dapps}.
This combination provides both high-fidelity spectral measurements and broad spatial coverage, feeding into the \gls{dt}'s interference predictions.
The monitoring data serves a dual purpose: in the short term, it triggers conflict detection when observed interference exceeds tenant thresholds; in the long term, it calibrates the \gls{dt}'s propagation models against ground truth, progressively improving the accuracy of predictive allocation decisions.

\begin{table*}[t!]
	\centering
	\caption{Per-Service simulation parameters, based on spectrum regulations from FCC, NTIA, and ITU, and technical specifications from relevant standard-development organizations. Note that all services overlap by at least 400 MHz in the spectrum between 3.1 GHz and 3.7 GHz, which is used as reference for the common pool in this paper.}
	\label{tab:service_params}
	\footnotesize
	\begin{tabular}{@{}llllllll@{}}
			\toprule
			\textbf{Service} & \textbf{Gain} & \textbf{Tx Pwr} & \textbf{Duty} & \textbf{Bandwidth} & \textbf{Typical Frequency Range} & \textbf{Nodes} & \textbf{Tx Height} \\
			\midrule
			Sensing         & 15--30~dBi  & 40--50~dBm  & 1--10\% & 200--400~MHz   & 3.1--3.7~GHz        & 10/50  & 30~m\\
			Radionavigation       & 20--28~dBi  & 60--75~dBm  & 0.1--1\% & 100--200~MHz  & 2.7--5.65~GHz       & 10/50 & 50~m\\
			Radiolocation       & 30--40~dBi  & 70--90~dBm  & 0.1--1\%  & 200--400~MHz  & 2.7--5.65~GHz       & 10/50 & 100~m\\
			Cellular        & 15--29~dBi  & 40--46~dBm  & 70--100\%  & 20--100~MHz & 2--4~GHz            & 50/100 & 25~m\\
			\bottomrule
	\end{tabular}
    \vspace{-4mm}
\end{table*}

\vspace{-3mm}
\section{\projName Operations}
\label{sec:workflows}

The \projName architecture enables two key operations for multi-service, multi-dimensional sharing. 

\vspace{-3mm}
\subsection{Conflict Detection and Resolution}
\vspace{-1mm}

The \gls{sic} continuously monitors the radio environment and compares observed conditions against each tenant's coexistence criteria.
When a conflict is detected---through predictive modeling (\gls{dt}) or real-time monitoring---the \gls{sic} initiates resolution on timescales of tens to hundreds of milliseconds.
Resolution strategies selected by the active \glspl{sApp} include temporal separation (duty-cycle adjustment), spatial separation (beam steering, power control), frequency reassignment, or priority-based preemption.
The \gls{sic}'s priority tables, configured per regulatory regime, encode service asymmetries: federal incumbents may receive preemptive protection, while commercial services negotiate shared access. This is enabled by the control that the \gls{sic} retains of the shared infrastructure pool: the architecture enables a direct sensing-to-enforcement loop.

As a concrete example, consider a weather radar (sensing) and a 5G base station sharing the 3.5~GHz band in the same geographic area.
If the radar begins a scan that would raise interference at the base station above the cellular tenant's coexistence threshold, the \gls{sic}'s duty-cycle enforcement \gls{sApp} first attempts temporal separation: it verifies whether the radar's low duty cycle ($\delta \approx 0.01$) leaves sufficient residual capacity for the cellular tenant and, if so, schedules non-overlapping slots.
If temporal separation is infeasible---e.g., because a third service already occupies the complementary slots---the \gls{sic} escalates to frequency reassignment, moving the cellular carrier to an adjacent channel.
Throughout, the \gls{dt} 
evaluates each candidate strategy before enforcement, reducing trial-and-error reconfiguration.

\vspace{-3mm}
\subsection{Service Sharing}
\vspace{-1mm}

\projName enables deploying non-cellular services on existing cellular infrastructure, as illustrated in Fig.~\ref{fig:service_sharing}.
Three modes are available: site sharing only (own \gls{rf} equipment at cellular sites), full \gls{rf}-chain sharing (adopting the host's antenna, power, and height parameters), and spectrum pooling with duty-cycle coordination.
The benefit depends on the match between native and host \gls{rf} parameters: as we quantify in Section~\ref{sec:eval}, services with cellular-like parameters see the greatest gains, while high-power services may experience degraded performance under site-sharing only, due to increased interference (unless managed).
The \gls{sic} selects among modes based on \gls{qos} requirements and the \gls{dt}'s compatibility assessment.

\vspace{-3.5mm}
\section{System-Level Evaluation}
\label{sec:eval}
\vspace{-0.5mm}

We validate the quantitative premises underlying the \projName architecture through system-level simulations based on the \gls{dt} component of the \gls{sic}, with the goal of characterizing the benefits and tradeoffs of infrastructure and spectrum sharing in Sections~\ref{sec:arch}--\ref{sec:workflows}. 
They do not constitute end-to-end validation of the \gls{sic}, \glspl{sApp}, or \gls{dt} feedback loop, which requires a full-stack prototype and is the subject of ongoing work.

\vspace{-3mm}
\subsection{Simulation Setup and Evaluation Scenarios}
\vspace{-0.5mm}

The simulation framework integrates the BostonTwin urban \gls{dt}~\cite{testolina2024bostontwin} with the Sionna ray-tracing engine~\cite{hoydis2023sionnart} for site-specific evaluation.
BostonTwin provides accurate 3D building geometry for a 2.3~km$^2$ urban tile, loaded into Sionna RT with radio material properties (permittivity, conductivity) assigned per building.
Antenna radiation patterns follow 3GPP TR~38.901 specifications.
Results are averaged over 50 Monte Carlo iterations with uniformly sampled user locations.
The deployment densities reflect the asymmetry between cellular and sparse services: cellular deploys $50$ base stations in 1~km$^2$, while each sparse service uses $10$ nodes in the baseline. 
This density asymmetry matters for the \textit{site-only sharing} case.
Sparse services that keep their native, higher transmit power while operating from the denser cellular grid generate far more aggregate interference than from the baseline deployment.
Therefore, in \textit{full-RF-chain} mode, their power is set to cellular levels.

\begin{figure}[t!]
	\centering
	\includegraphics[width=0.75\columnwidth]{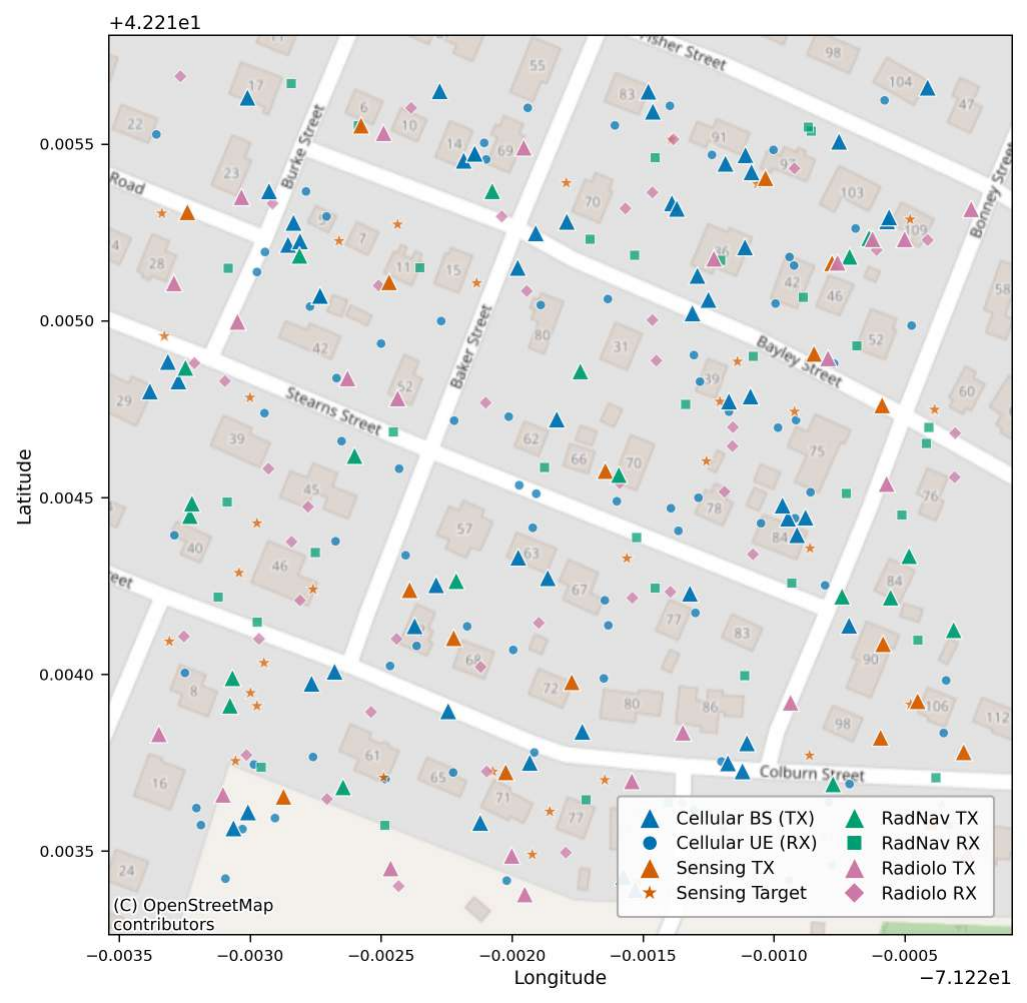}
	\caption{Multi-service deployment in BostonTwin. Triangles: base stations; dots: user equipment. Colors distinguish services. 
    }
	\label{fig:netw_deploy_sionna}
    \vspace{-4mm}
\end{figure}

The simulation captures inter-service interference by computing aggregate received power from all co-channel transmitters, including cross-service contributions (e.g., cellular base stations interfering with radar receivers and vice versa), providing a realistic assessment of coexistence tradeoffs. 
The \gls{sinr} is modeled per service: cellular downlink follows the standard linear receiver; active sensing uses the two-way radar equation~\cite{skolnik2001introduction}; radionavigation and radiolocation report post-correlation \gls{sinr} at the strongest-beacon receiver, with cross-service transmitters treated as interference throughout.

Table~\ref{tab:service_params} summarizes the per-service parameters (sourced from U.S. and international spectrum policy regulations), while Fig.~\ref{fig:netw_deploy_sionna} shows a realization of the multi-service deployment.
We consider a 400\,MHz shared spectrum pool centered around \textcolor{black}{3.5}\,GHz and a thermal noise floor of $-174$~dBm/Hz.
%



We evaluate three configurations. In~\textbf{Baseline}, each service operates on its own exclusive band with dedicated infrastructure (current state of the art); 
specifically, services are centered at $f_c = 3.5$~GHz (cellular), $3.4$~GHz (sensing), $2.8$~GHz (radionavigation), and $3.3$~GHz (radiolocation), so cross-service 	interference is absent by construction.
With \textbf{Infrastructure sharing}, a sparse service reuses cellular deployment 
at $f_c = 3.5$~GHz,
either adopting cellular \gls{rf} parameters (\textit{full-RF-chain sharing}) or deploying its own equipment at cellular sites (\textit{site-only sharing}). We assume that, in the case of sharing, the \gls{sic} can coordinate access without generating harmful \gls{rfi} across services.

\subsection{SINR Performance under Infrastructure Sharing}

    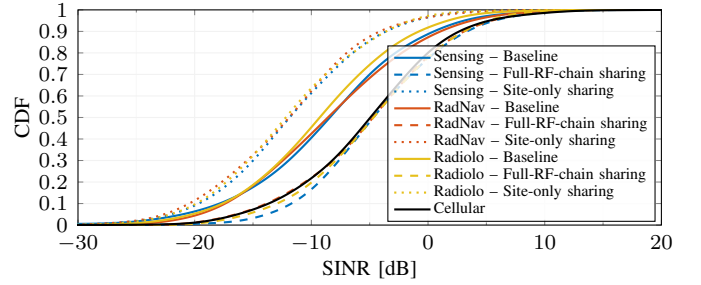
\begin{figure}[t!]
		\centering
		\begin{tikzpicture}
			\begin{axis}[
				width=1.05\linewidth, 
				height=0.5\linewidth, 
				xlabel={\scriptsize SINR [dB]},
				ylabel={\scriptsize CDF},
				tick label style={font=\scriptsize},
				grid=both,
				major grid style={gray!10},
				minor grid style={gray!5},
				major tick length=2pt,
				minor tick length=1pt,
				ymin=0, ymax=1,
				xmin=-30, xmax=20,
				xtick distance=10,
				ytick distance=0.1,
				minor x tick num=1,
				label shift=-4pt,
				legend cell align=left,
				legend style={
					at={(0.99,0.015)}, anchor=south east,
					font=\tiny,
					fill=white, fill opacity=0.8, draw opacity=1, text opacity=1,
					inner xsep=2pt, inner ysep=0.8pt,
					nodes={inner sep=0.8pt},
					row sep=-1pt,
				},
                legend image post style={xscale=0.8},
				]
				
				\addplot[cdfcurve, color=svcSensing,  solid,  mark=none]
				table[x=sinr_dB, y=cdf] {cdf_data_all/cdf_Sensing_Baseline.dat};
				\addlegendentry{Sensing -- Baseline}
				
				\addplot[cdfcurve, color=svcSensing,  dashed, mark=none]
				table[x=sinr_dB, y=cdf] {cdf_data_all/cdf_Sensing_InfraFull.dat};
				\addlegendentry{Sensing -- Full-RF-chain sharing}
				
				\addplot[cdfcurve, color=svcSensing,  dotted, mark=none]
				table[x=sinr_dB, y=cdf] {cdf_data_all/cdf_Sensing_InfraSite.dat};
				\addlegendentry{Sensing -- Site-only sharing}
				
				\addplot[cdfcurve, color=svcRadNav,   solid,  mark=none]
				table[x=sinr_dB, y=cdf] {cdf_data_all/cdf_RadNav_Baseline.dat};
				\addlegendentry{RadNav -- Baseline}
				
				\addplot[cdfcurve, color=svcRadNav,   dashed, mark=none]
				table[x=sinr_dB, y=cdf] {cdf_data_all/cdf_RadNav_InfraFull.dat};
				\addlegendentry{RadNav -- Full-RF-chain sharing}
				
				\addplot[cdfcurve, color=svcRadNav,   dotted, mark=none]
				table[x=sinr_dB, y=cdf] {cdf_data_all/cdf_RadNav_InfraSite.dat};
				\addlegendentry{RadNav -- Site-only sharing}
				
				\addplot[cdfcurve, color=svcRadiolo,  solid,  mark=none]
				table[x=sinr_dB, y=cdf] {cdf_data_all/cdf_Radiolo_Baseline.dat};
				\addlegendentry{Radiolo -- Baseline}
				
				\addplot[cdfcurve, color=svcRadiolo,  dashed, mark=none]
				table[x=sinr_dB, y=cdf] {cdf_data_all/cdf_Radiolo_InfraFull.dat};
				\addlegendentry{Radiolo -- Full-RF-chain sharing}
				
				\addplot[cdfcurve, color=svcRadiolo,  dotted, mark=none]
				table[x=sinr_dB, y=cdf] {cdf_data_all/cdf_Radiolo_InfraSite.dat};
				\addlegendentry{Radiolo -- Site-only sharing}
				
				\addplot[cdfcurve, color=svcCellular, solid,  mark=none]
				table[x=sinr_dB, y=cdf] {cdf_data_all/cdf_Cellular_Baseline.dat};
				\addlegendentry{Cellular}
				
			\end{axis}
		\end{tikzpicture}
		\caption{CDF of downlink \gls{sinr}: baseline (solid), full-RF-chain sharing (dashed), site-only sharing (dotted). Cellular in black. 
        }
		\label{fig:cdf_all_si}
        \vspace{-5mm}
	\end{figure}

Figure~\ref{fig:cdf_all_si} presents the CDFs of downlink \gls{sinr} under three infrastructure configurations.
The baseline configuration groups the three CDFs for sensing, radionavigation (RadNav), and radiolocation (Radiolo) around medians within $-9$ and $-8$ dB.
Sensing shows higher SINR (i.e., the power over noise and interference at target) for farther away targets compared to radio services for navigation and location. Under site-only sharing, the higher transmit power for sensing, location, and navigation when compared to cellular leads to increased interference when the services are deployed at the much higher density presented by the cellular network.
Full-RF-chain sharing, by contrast, removes the interference penalty.
The combination of increased density and fine-tuned RF parameters (including transmit power) lead to an alignment between the cellular \gls{sinr} and the other services, with a median improvement of $12$ dB.  

These results quantify the incentive that the sensing, radiolocation, and radionavigation spectrum services may have in joining the common pool: access to high-density infrastructure can lead to improved \gls{sinr}, as long as \gls{rf} parameters are properly tuned. 
The \gls{dt} enables such tuning \textit{before} deployment: by simulating each candidate configuration against BostonTwin's 3D geometry, the \gls{sic} can predict which mode will meet a tenant's \gls{qos} target and avoid costly reconfiguration.

It is worth noting the asymmetry between the uplink and downlink perspectives.
The results presented here focus on downlink \gls{sinr} at user/receiver locations, but the interference landscape differs for uplink-dominated services such as passive sensing, where interference is experienced at the base station or sensor rather than at a distributed user population.
Passive receivers cannot adjust their transmit power, so their coexistence relies entirely on the \gls{sic} scheduling active transmitters around passive observation windows---well-suited to the duty-cycle enforcement \gls{sApp}. 

\vspace{-2mm}
\subsection{Effective Rate vs.\ Duty Cycle and Bandwidth}

    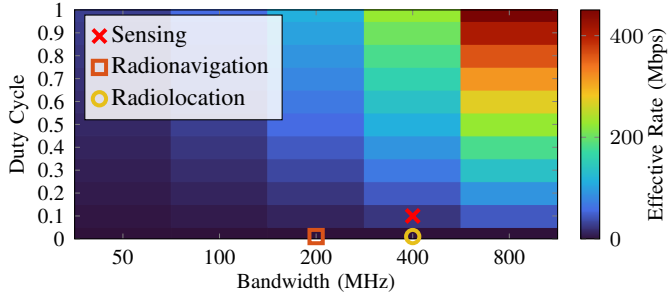
\begin{figure}[t!]
		\centering
		\begin{tikzpicture}
			\begin{axis}[
				width=0.9\linewidth, 
				height=0.52\linewidth,
				enlargelimits=false,
				axis on top,
				xlabel={\scriptsize Bandwidth (MHz)},
				ylabel={\scriptsize Duty Cycle},
				xtick={1,2,3,4,5},
				xticklabels={50,100,200,400,800},
				ytick={0,0.1,0.2,0.3,0.4,0.5,0.6,0.7,0.8,0.9,1},
				yticklabels={0,0.1,0.2,0.3,0.4,0.5,0.6,0.7,0.8,0.9,1},
				xmin=0.5, xmax=5.5,
				ymin=0, ymax=1,
				point meta min=0,
				point meta max=450.677,
				tick label style={font=\scriptsize},
				major tick length=2pt,
				label shift=-4pt,
				colormap name=turbo,
				colorbar,
				colorbar style={
					width=0.25cm,
					ylabel={\scriptsize Effective Rate (Mbps)},
					ylabel shift=-6pt,
					major tick length=2pt,
					minor tick length=1pt,
				},
				legend style={
					at={(0.02,0.98)}, anchor=north west,
					fill=white, fill opacity=0.85,
					draw opacity=1, text opacity=1,
					font=\footnotesize,
				},
				legend cell align=left,
				grid=none,
				tick align=inside,
				]
				
				\addplot[
				matrix plot*,
				mesh/cols=5,
				mesh/ordering=y varies,
				point meta=explicit,
				forget plot,
				] table[x=bw_idx, y=delta, meta=rate] {rate_heatmap.dat};
				
				\addplot+[only marks, mark=x, mark size=3.5pt, color=red, mark options={line width=1.5pt, fill=none}] coordinates {(4,0.1)};
				\addlegendentry{Sensing}
				
				\addplot+[only marks, mark=square, mark size=2.5pt, color=svcRadNav, mark options={line width=1.5pt, fill=none}] coordinates {(3,0.01)};
				\addlegendentry{Radionavigation}
				
				\addplot+[only marks, mark=o, mark size=2.5pt, color=svcRadiolo, mark options={line width=1.5pt, fill=none}] coordinates {(4,0.01)};
				\addlegendentry{Radiolocation}
				
			\end{axis}
		\end{tikzpicture}
		\caption{Cellular effective rate vs.\ duty cycle and bandwidth. Markers: sensing ($\times$), radionavigation ($\square$), radiolocation ($\circ$).}
		\label{fig:rate_duty_bw_all_s0}
        \vspace{-5mm}
	\end{figure}


\textcolor{black}{
Figure~\ref{fig:rate_duty_bw_all_s0} characterizes the cellular effective rate, defined as $R_{\mathrm{eff}} = \delta \cdot B \cdot \overline{C}$, where $\overline{C}$ is the median spectral efficiency from the ray-traced \gls{sinr} distribution, as a function of duty cycle $\delta$ and bandwidth $B$, when sparse services share the same infrastructure as cellular under \gls{sic}-coordinated non-overlapping scheduling.
}

\textcolor{black}{The duty-cycle ranges in Table~\ref{tab:service_params} are sourced from NTIA radar characterizations~\cite{ntia_3100_3700} and spectrum regulations~\cite{SS:UpperMid24},}
and the red markers indicate the nominal operating points of each sparse service: sensing ($\times$) at \textcolor{black}{$\delta \sim 0.1$} and \textcolor{black}{$B\sim400$} MHz, radionavigation ($\square$) at \textcolor{black}{$\delta \sim 0.01$} and \textcolor{black}{$B\sim210$} MHz, and radiolocation ($\circ$) at \textcolor{black}{$\delta \sim 0.01$} and \textcolor{black}{$B\sim400$} MHz.
\textcolor{black}{These operating points sit well below the cellular regime ($\delta \sim 0.9$--$1.0$), confirming that sparse services typically use less than $10\%$, and often less than $1\%$, of available time-frequency resources.}
When the \gls{sic} can schedule non-overlapping time slots (i.e., $\delta_X + \delta_Y \leq 1$ with synchronized slot boundaries and guard intervals), spectrum sharing is interference-free.
When duty-cycle budgets are incompatible or services have inflexible timing requirements (e.g., radars with fixed dwell schedules), the \gls{sic} resorts to frequency-domain or spatial separation.
With four or more co-channel services, the generalized constraint $\sum_i \delta_i \leq 1$ makes scheduling combinatorial, requiring the \gls{sic}'s automated coordination.

These rate curves also quantify the economic case for spectrum sharing.
A sensing service operating at \textcolor{black}{ $\delta = 0.1$ leaves $90\%$ of time-frequency resources unused; a radionavigation service at $\delta = 0.01$ leaves $99\%$.}
Under \projName, the \gls{sic} can allocate these idle slots to other tenants, increasing aggregate spectral efficiency without degrading existing services. 
The surface in Fig.~\ref{fig:rate_duty_bw_all_s0} further shows diminishing returns at wider bandwidths for low-duty-cycle services, suggesting regulators could specify maximum bandwidth--duty-cycle products rather than fixed band allocations.

\vspace{-4.5mm}
\section{Discussion}
\label{sec:discussion}
\vspace{-1.5mm}

The evaluation results reveal several insights that inform both the \projName architecture and broader spectrum policy.

\textbf{Infrastructure sharing is not one-size-fits-all.}
The different nature of incentives and performance gains for heterogeneous services points to the \gls{sic} operating as a matchmaker, not a simple resource allocator.
Each sharing request requires a compatibility assessment that considers the full parameter vector---power, gain, duty cycle, antenna pattern, location, and height---rather than frequency alone.
This contrasts with current \gls{sas} designs, which primarily manage frequency-domain access.

\textbf{Temporal complementarity is the key enabler.}
The duty-cycle analysis shows that most non-cellular services use less than $1$--$10\%$ of available time-frequency resources.
This temporal sparsity is the primary resource that \projName exploits: by coordinating time-domain access across services, the \gls{sic} can achieve interference-free coexistence without the power reductions or geographic exclusion zones that characterize current sharing frameworks.

\textbf{Hardware compatibility constrains sharing depth.} Full-\gls{rf}-chain sharing assumes that sparse services can adopt cellular \gls{rf} parameters, but radar waveforms require high \gls{papr} and amplifiers exceeding the ${\sim}46$~dBm ceiling of cellular radio units, while navigation beacons use narrowband waveforms that may conflict with the \gls{nr} subcarrier spacing. Full-\gls{rf}-chain sharing is thus currently viable only for sensing-like services, whereas high-power services should begin with site-only sharing and transition as reconfigurable \gls{sdr} front-ends mature.

\textbf{Economic viability requires cellular and passive sensing \gls{qos} preservation.}
Cellular operators will only participate in \projName if hosting additional services does not trigger \gls{qos} violations or customer complaints.
The tight interaction between \gls{sic} and shared infrastructure is a key enabler for both.

\textbf{\glspl{dt} bridge the gap between conservative and aggressive sharing.}
Current exclusion zones are deliberately conservative, protecting incumbents with large geographic margins that leave spectrum unused.
Site-specific propagation modeling can tighten these margins by accounting for actual building geometry and terrain, but this requires validated models and continuous calibration---capabilities that the \projName monitoring framework is designed to provide.
The transition from static exclusion zones to dynamic, model-driven sharing boundaries represents a gradual deployment path rather than a binary switch.

\vspace{-3mm}
\section{Conclusions}
\label{sec:conclusions}

This article introduced \projName, an architecture that goes beyond current \gls{dsa} frameworks by jointly sharing spectrum, services, and infrastructure.
The \gls{sic} with plug-and-play \glspl{sApp}, shared infrastructure pool, and \gls{dt}-based coordination address key limitations of existing approaches: pairwise-only coexistence, binary grant/deny decisions, and spectrum-only sharing.
System-level simulations 
show the impact of infrastructure sharing incentives and that duty-cycle complementarity enables interference-free coexistence under \gls{sic} coordination, with cellular \gls{qos} preserved.

This article serves as the foundational introduction for the \projName concept. The building blocks---\gls{oran}, \glspl{dt}, neutral hosting---exist today; realizing the \projName vision requires their integration alongside new standardization and policy efforts.
Future work includes prototyping the system and evaluating scalability, resource allocation, security, and techno-economic implications. 
On the regulatory side, future frameworks should adopt service-agnostic grant structures specifying interference constraints rather than technology-specific operating rules.


\vspace{-2mm}
\footnotesize{
\bibliographystyle{IEEEtran}
\bibliography{ref_26}
}

\vspace{-1.5cm}
\begin{IEEEbiographynophoto}
\noindent\textbf{Michele Polese} is a Research Assistant Professor at Northeastern University. He received his Ph.D. from the University of Padova in 2020.

\noindent\textbf{Minh Dat Nguyen} is a Postdoctoral Researcher at Northeastern University. He received his Ph.D. from the Institut National de la Recherche Scientifique (INRS) in 2023.

\noindent\textbf{Paolo Testolina} is a Research Scientist at Northeastern University. He received his Ph.D. from the University of Padova in 2023.

\noindent\textbf{Tommaso Melodia} is the William Lincoln Smith Chair Professor at Northeastern University and Director of the Institute for Intelligent Networked Systems.
\end{IEEEbiographynophoto}

\end{document}